**Absentee and Economic Impact of Low-Level Fine Particulate Matter and Ozone Exposure in K-12 Students**

Daniel L. Mendoza[1,2], Cheryl S. Pirozzi[1], Erik T. Crosman[3], Theodore G. Liou[1,4], Yue Zhang[5], Jessica J. Cleeves[6], Stephen C. Bannister[7], William R.L. Anderegg[8], Robert Paine III[1]

[1] Division of Respiratory, Critical Care and Occupational Pulmonary Medicine, School of Medicine, University of Utah

[2] Department of Atmospheric Sciences, University of Utah

[3] Department of Life, Earth, and Environmental Sciences, West Texas A&M University

[4] Center for Quantitative Biology, University of Utah

[5] Division of Epidemiology, Department of Internal Medicine, University of Utah School of Medicine

[6] Center for Science and Mathematics Education, University of Utah

[7] Department of Economics, University of Utah

[8] School of Biological Sciences, University of Utah

Corresponding author: Daniel L. Mendoza; Division of Respiratory, Critical Care and Occupational Pulmonary Medicine, Health Sciences Center, 26 North 1900 East, Salt Lake City, Utah 84132; daniel.mendoza@utah.edu; (765) 430-7367

**Abstract**

High air pollution levels are associated with school absences. However, low level pollution impact on individual school absences are under-studied. Positive local findings could improve school recess decisions, better identify pollution sources and improve local economic effects assessments. We modelled $PM_{2.5}$ and ozone concentrations at 36 schools from July 2015 to June 2018 using data from a dense, research grade regulatory sensor network. We determined exposures and daily absences at each school. We used generalized estimating equations model to retrospectively estimate rate ratios for association between outdoor pollutant concentrations and school absences. We estimated lost school revenue, productivity, and family economic burden. $PM_{2.5}$ and ozone concentrations and absence rates vary across the School District. Pollution exposure were associated with as high a rate ratio of 1.02 absences per $\mu g/m^3$ and 1.01 per ppb increase for $PM_{2.5}$ and ozone, respectively. Significantly, even $PM_{2.5}$ and ozone exposure below regulatory standards ($< 12.1\ \mu g/m^3$ and $< 55$ ppb) was associated with positive rate ratios of absences: 1.04 per $\mu g/m^3$ and 1.01 per ppb increase, respectively. Granular local measurements enabled demonstration of air pollution impacts that varied between schools undetectable with averaged pollution levels. Reducing pollution by 50% would save $452,000 per year districtwide. Pollution reduction benefits would be greatest in schools located in socioeconomically disadvantaged areas. Exposures to air pollution, even at low levels, are associated with increased school absences. Heterogeneity in exposure, disproportionately affecting socioeconomically disadvantaged schools, points to the need for fine resolution exposure estimation. The economic cost of absences associated with air pollution is substantial even excluding indirect costs such as hospital visits and medication. These findings may help inform

decisions about recess during severe pollution events and regulatory considerations for localized pollution sources.

**Keywords**

# 1. Introduction

Exposure to air pollution worsens health by increasing hospitalizations [1-2] due to cardiovascular and pulmonary events [3-4], asthma exacerbations [5], and mortality [6-7]. Ozone and fine particulate matter ($PM_{2.5}$) are prominent pollutants associated with negative health outcomes with a recent study demonstrating the effects of even low levels of exposure on mortality [8]. Measures to curb these emissions have improved health outcomes [9]. Children are an especially vulnerable population due to their higher ventilatory rates, level of activity, and time spent outdoors that increases their exposure to air pollution. Several studies have focused on the impact of environmental hazards on children, with specific additional emphasis on environmental justice [10-14].

One potential adverse health effect from air pollution is increased school absence days. Unfortunately, elevated levels of multiple pollutants including $PM_{2.5}$, coarse particulate matter ($PM_{10}$), ozone, nitrogen dioxide ($NO_2$), and carbon monoxide (CO) are common near schools [15], likely due to both school placement and transportation-related emissions. Elevated pollution, including $PM_{10}$, ozone, and oxides of nitrogen ($NO_x$), contributes to school absences [16-19] even at low levels [20].

Chronic absenteeism in elementary and middle school has long-term implications, reliably predicting failure to graduate high school and lower individual lifetime earnings. High-school drop-outs earn $10,386 per year less than their diploma-holding counter-parts, and are twice as likely to live in poverty than college graduates [21-22]. School absenteeism exacerbates social class differences in academic development, and higher attendance rates benefit lower socioeconomic status children the most [23].

Therefore, environmental factors that may contribute to school absenteeism may have important long-term societal consequences.

Because the burden of poor air quality is not shared equally among populations [24-25] it is critically important to study environmental exposure at neighborhood scales. In this study, we use a dense pollutant observation platform to provide high quality estimations of $PM_{2.5}$ and ozone exposures at individual schools in the Salt Lake City School District (SLCSD). Using this granular data, we analyzed the effect of low air pollution time periods on school absences. We estimated the economic impact of absences associated with air pollution as costs to individual schools, families, and the overall economy.

**2. Methods**

2.1 Air Pollution Exposure Modeling

The Salt Lake City Metropolitan area is home to a dense criteria air pollution observational network facilitating a wide range of observation studies [26-28]. We combine data from the Utah Division of Air Quality (UDAQ) regulatory observation network, and the University of Utah (UofU) stationary and mobile platform network. The mobile network consists of air quality sensors that measure $PM_{2.5}$ and ozone [29-30] mounted on top of electric Utah Transit Authority (UTA) light-rail trains (Appendix A, Fig. A.1). Using $PM_{2.5}$ and ozone data over three years (July 2015 to June 2018), we estimated air pollution exposure at each school at 5-minute resolution using an inverse distance square weighting (IDW) method [31-32]. Since only two schools are located farther than 4km from a sensor, IDW was an appropriate exposure estimation method.

2.2 Absences Data

The Salt Lake City School District (SLCSD) is entirely within the boundaries of Salt Lake City, Utah (Appendix A, Fig. A.1). Salt Lake City has a sociodemographic West-East division, with the west side home to a higher proportion of lower-income (Title 1 schools) and minority communities than the east side (Appendix A, Table A.1). Of the 36 schools in the SLCSD (20 east, 16 west), 26 are elementary schools (14 east, 12 west), 7 are middle schools (4 east, 3 west), and 3 are high schools (2 east, 1 west). Full-day daily absences for each school were provided for this study by the SLCSD.

2.3 Study Measures

We estimated exposure during school hours, recess hours, and the daily average for the specific day, ranging from one to five days lag. We selected 7 AM to 3 PM to represent "school day hours", and 10 AM to 2 PM to represent "recess hours" to encompass the time students could be outdoors. The temporal metrics were grouped by either the full academic year or season temporal scale: Fall (September, October, November), Winter (December, January, February), and Spring (March, April, May). The Summer had too few school days to be included in this study. Results were grouped by grade level (elementary, middle, high school) and also by west and east side schools, as well as district wide. Pollutant exposure was categorized for all levels or low-levels, defined by "good" air quality (< 12.1 µg/m$^3$ for $PM_{2.5}$; < 55 ppb for ozone), according to the United States Environmental Protection Agency's (USEPA) Air Quality Index (AQI) Recommendations [33].

2.4 Statistical Procedures

We used Generalized Estimating Equation models (R and the package 'geepack' Version 1.2-1) [34-37] with independence working correlation structures to estimate the association of $PM_{2.5}$ and ozone exposure, at the individual school level, as the independent variables with school absences as the dependent variable, with adjustment for plausible confounding variables, including temperature at 7 AM [38], pollen counts [39], and influenza-related hospitalizations and hospital visits [40]. Rate ratios and 95% confidence intervals (CI) were calculated for each school for absences per unit of pollutant exposure: $\mu g/m^3$ of for $PM_{2.5}$ or parts per billion for ozone.

2.5 Economic Analysis

Negative outcomes ("externalities") of exposure to air pollution can be quantified and assessed for economic impact to help prioritize policy responses. We performed a case study involving a reduction of 24-hour exposure to $PM_{2.5}$ and ozone by 50% from the current values to estimate the potential reduction in absences associated with exposure reduction.

Utah median expenditure for non-charter and mixed school districts is approximately $7,434 per pupil annually, or $41.30 per pupil per day in a 180 school-day year [41]. The number of absences were multiplied by $41.30 to obtain the potential revenue increase for each school. The current subsidized costs of breakfast and lunch are $3.00, $3.50, and $3.70 for elementary, middle, and high school students, respectively. These costs were multiplied by 0.56 (the proportion of SLCSD students eligible for free or reduced cost meals) to estimate the family burden of food costs since an absent student would not receive these meals at school.

The average hourly wage in 2019 dollars was estimated to be $23.74 per hour [42-43]. This value was used to calculate the lost wages (8-hour workday) from a parent staying at home to care for an absent child. Lost wages are only a fraction of lost economic productivity. A conservative economic multiplier estimate is 2.5 [44] which accounts for externalities including lost revenue, lost taxes, and lost worker output. This value was multiplied by the value derived from the lost wages calculation to

estimate the total lost economic productivity. The full economic impact of absences was derived as the sum of lost school revenue, family burden of food costs, lost wages, and total lost economic productivity.

The complete pollution observation and absences data set was available for the entire study period, thus there is no missing data for this analysis.

This study was determined to be exempt by the IRB of the University of Utah.

**3. Results**

3.1 Pollutant exposure, absences, and geographical variables

Geography and urban development patterns influence pollution levels within the Salt Lake Valley. The west side of Salt Lake City has more emission sources contributing to $PM_{2.5}$ levels than the east side, including the airport, freight railroads, congested highways, and industrial facilities. $PM_{2.5}$ exposure decreases west to east based on the concentration of sources and lower elevation on the west side (Fig. 1A, Appendix A, Fig. A.2B). Conversely, ozone concentrations decrease east to west, the opposite geographical trends of $PM_{2.5}$ (Fig. 1B, Appendix A, Fig. A.2D), due to elevation effects and titration of ozone by nitrogen oxides at the lower elevations [27]. The western and lower elevation areas show higher absence rates (Fig. 1C, Appendix A, Fig. A.2F). In contrast, there is no clear association between latitude and either pollution exposure or school absences (Appendix A, Fig. A.2A,C,E).

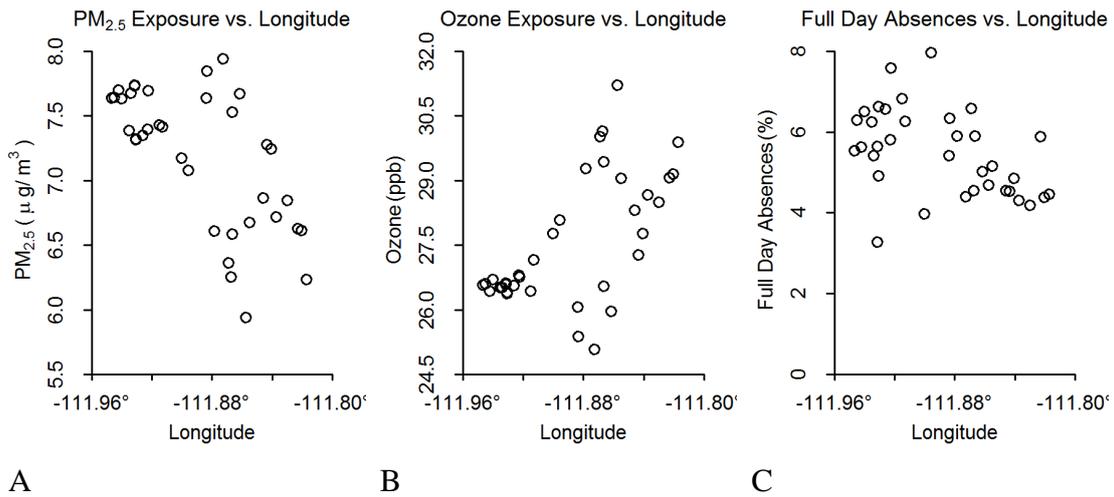

A B C

Fig. 1. Pollutant exposure and absences by longitude

Comparison between average annual pollutant exposure and absences by longitude. Each circle represents a school location and its corresponding annual pollution exposure or absence rate: A) fine particulate matter ($PM_{2.5}$) vs. longitude, B) ozone vs. longitude, and C) absences vs. longitude.

Although the annual average values are low by National Ambient Air Quality Standards (NAAQS), episodic high levels of pollutants are a large concern in the Salt Lake City Metropolitan area [45]. The monthly average exposure for December 2017, a typical winter month, for all schools is above the yellow (or "moderate") level (12.1 – 35.4 µg m$^{-3}$), with some schools on the west side displaying up to 6 µg m$^{-3}$ greater average $PM_{2.5}$ levels compared to east side schools (Fig. 2A).

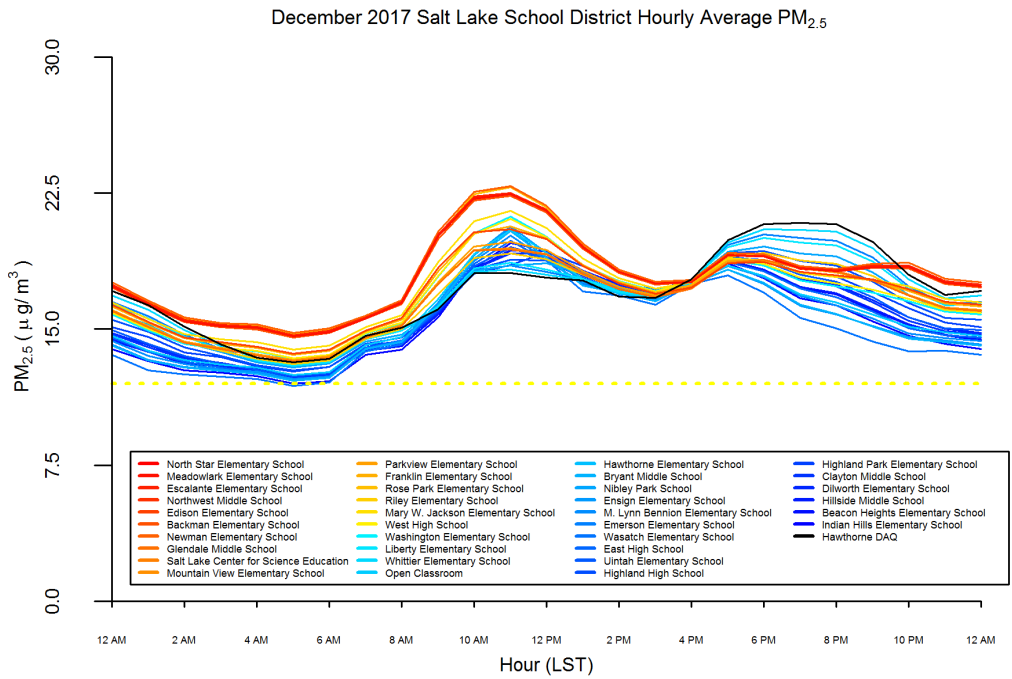

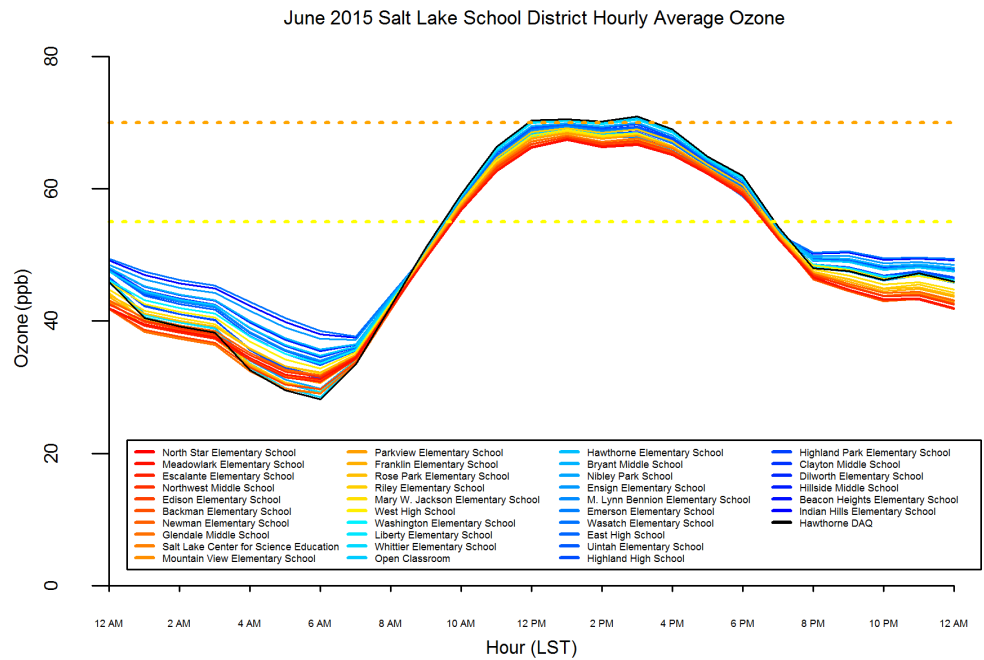

Fig. 2. Monthly hourly pollutant concentration averages

Diurnal patterns of A) fine particulate matter ($PM_{2.5}$) and B) ozone across the Salt Lake City School District. The red colors represent schools farther west on the district, and blue represent schools farther east. The black line represents readings at the Division of Air Quality regulatory sensor. Dashed horizontal lines represent air quality index (AQI) levels.

Ozone levels are high in the afternoon and lower at night, due to photochemical reactions during the day producing a rapid increase in ozone [27] (Fig. 2B). Schools on the eastern part of the SLCSD have consistently, but modestly, higher levels of ozone, the opposite pattern from $PM_{2.5}$. During daytime hours, critically during school and recess hours, ozone levels are generally in the yellow level (55 – 70 ppb) and rise to the orange ("unhealthy for sensitive groups") levels (71 – 85 ppb) at some schools on the east side. However, at night, ozone levels drop to below the yellow level to green ("good") levels (< 55 ppb) for all schools.

3.2 Pollutant exposure and absences

Like exposure rates, the background absence rates vary across SLCSD schools (Table 1, Appendix A, Table A.1). However, prior day $PM_{2.5}$ exposure is associated with increased school absences at all grade levels in both east and west side schools in the Fall (Fig. 3). The rate ratios are similar for all grade levels and geographical location, despite differences in local $PM_{2.5}$ levels and differences in background absence rate. The school day (Appendix A, Fig. 3) and recess (Appendix A, Fig. 4) exposure and absence rate ratios show similar patterns for similar time periods. The strongest associations for both time periods are with the average 24-hour exposure.

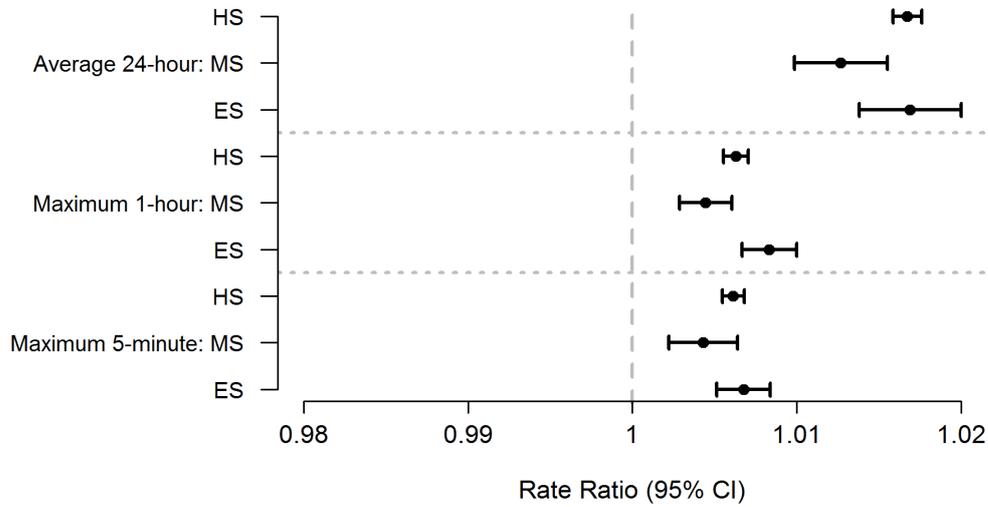

A

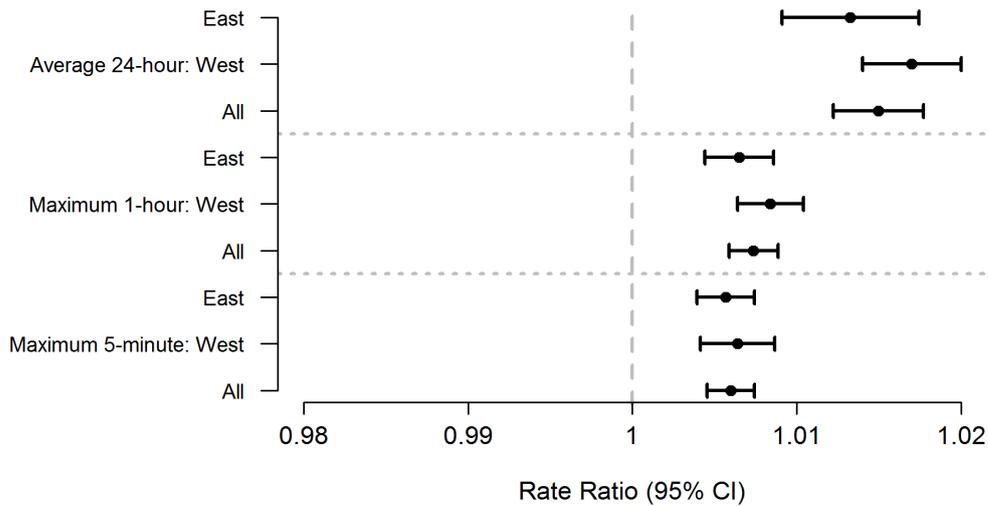

B

Fig. 3. Association between absences and previous day fine particulate matter (PM$_{2.5}$) exposure during the fall

Rate ratios of fall (September-November) absences associated with previous day fine particulate matter (PM$_{2.5}$) exposure for: A) all school levels where "ES", "MS", and "HS" correspond to elementary, middle, and high school, respectively, and B) elementary schools disaggregated by geographical location: "All" is all schools, "West" is all west side schools, and "East" is all east side schools.

For all exposure levels, the absence rate ratio is generally highest (~1.02) for 1-day lag after the PM$_{2.5}$ exposure measurements for all school levels (Fig. 4A). Even low levels of PM$_{2.5}$ exposure (<12.1 µg/m$^3$) level exposure result in increased absences, with rate ratios as high as 1.04 (Fig. 4B). Elementary school children have rate ratios associated with low level exposure that are higher than for older children. These findings are similar for the spring season.

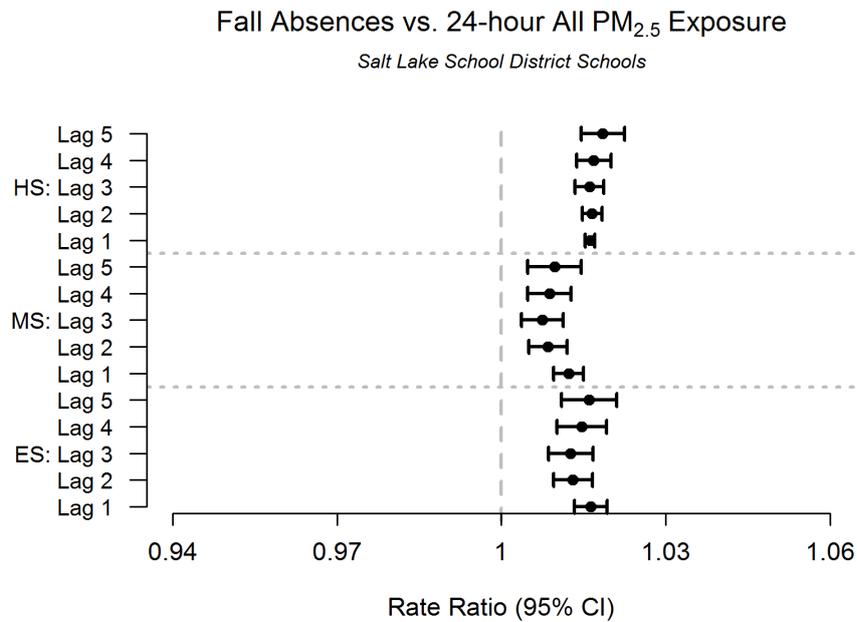

A

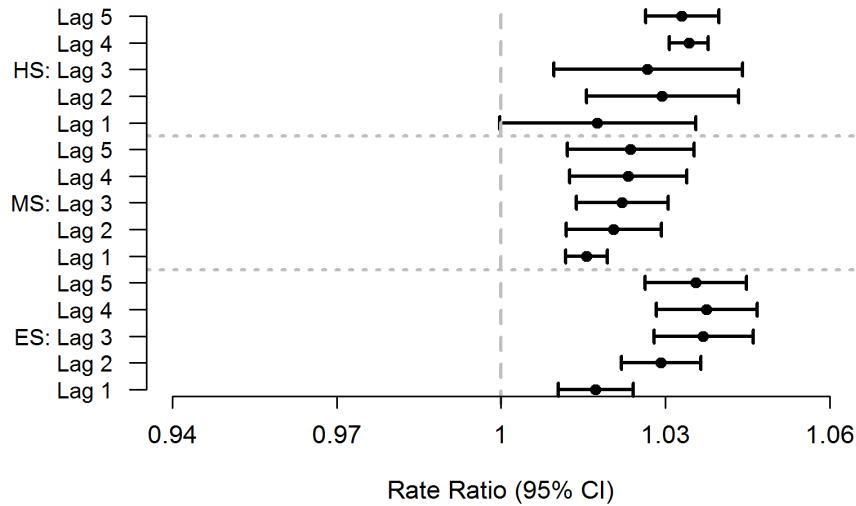

B

Fig. 4. Association between absences and lagged fine particulate matter (PM$_{2.5}$) exposure during the fall

Rate ratios of Salt Lake City School District fall (September-November) absences associated with lagged fine particulate matter (PM$_{2.5}$) exposure for: A) all concentration levels and B) low (<12.1 µg/m$^3$) concentration levels.

There is a similar association between PM$_{2.5}$ and school absences in the winter. However, because we found such a strong correlation between PM$_{2.5}$ concentrations and temperature during winter cold air pool events [46], it was impossible to differentiate between the impact of temperature and PM$_{2.5}$.

Absence rate ratios show a positive relationship with prior day ozone exposure across the full academic year, Winter, Spring, and the majority of temporal and school level metrics Appendix A, Figs. A.5-7, with rate ratios as high as 1.01 per ppb increase for both all and low (<55 ppb) exposure levels.

3.3 Economic benefit of reductions in school absences

The estimated potential reductions in absences achievable by reducing concentration levels of $PM_{2.5}$ and ozone by half range from 437 per year for all elementary schools to 50 per year for all high schools across the school district (Table 1). The associated economic impact of these reduced absences in the SLCSD would be approximately $452k per year (Table 1, Appendix A, Table A.1). Appendix A, Table A.1 disaggregates the results by school location instead of grade level. Schools on the west side of the SLCSD teach students of lower socioeconomic and higher minority background on average, and the majority of Title 1 schools (schools with a student base that are lower-income) [47] are located on the west side. Schools on the west side of the SLCSD have a higher rate of baseline absenteeism (5.82% vs. 5.40%) and, despite serving a markedly smaller total student population (10,368 vs. 12,380). If all exposure rates were reduced in half west side schools would experience a proportionately larger reduction in annual absences (31.15 vs. 25.52 per thousand students) than east side schools. This larger reduction is related to the higher average levels of $PM_{2.5}$ at west side schools.

Table 1. Economic impact of school absences in the Salt Lake City School District (SLCSD) and potential benefits of pollution reduction disaggregated by school level.

| School Level (N) | Elementary (26) | Middle (7) | High (3) | District (36) |
|---|---|---|---|---|
| **Title 1 Schools** | 16 | 3 | 0 | 19 |
| **Annual Enrollment: Mean (SD)** | 486 (104) | 586 (199) | 2,002 (454) | 632 (451) |
| **Annual Attendance Days: Mean (SD)** | 87,515 (18,787) | 105,471 (35,768) | 360,344 (81,673) | 113,742 (81,177) |
| **Annual Absence Days: Mean (SD)** | 4,951 (1,147) | 5,623 (2,284) | 16,051 (2,617) | 6,007 (3,425) |
| **Absences (%): Mean (SD)** | 5.78 (1.19) | 5.34 (1.22) | 4.56 (0.59) | 5.59 (1.19) |
| **Annual PM$_{2.5}$ (μg/m$^3$): Mean (SD)** | 6.61 (0.36) | 6.43 (0.34) | 6.5 (0.24) | 6.56 (0.35) |
| **Annual Ozone (ppb): Mean (SD)** | 28.34 (0.74) | 28.73 (0.59) | 28.61 (0.45) | 28.44 (0.7) |
| **Absences Reduction: Mean (SD)** | 17 (9) | 22 (10) | 17 (5) | 18 (9) |
| **Lost School Revenue ($): Mean (SD)** | 695 (366) | 898 (417) | 684 (212) | 733 (367) |
| **Lost Wages ($): Mean (SD)** | 3,195 (1,684) | 4,127 (1,916) | 3,144 (973) | 3,372 (1,689) |
| **Meal Costs ($): Mean (SD)** | 28 (15) | 43 (20) | 34 (11) | 32 (16) |
| **Total Family Burden ($): Mean (SD)** | 3,223 (1,699) | 4,170 (1,935) | 3,179 (984) | 3,404 (1,705) |
| **Economic Multiplier ($): Mean (SD)** | 7,987 (4,211) | 10,318 (4,789) | 7,861 (2,433) | 8,430 (4,221) |
| **Total Economic Cost: Mean (SD)** | 11,905 (6,276) | 15,386 (7,141) | 11,724 (3,629) | 12,567 (6,293) |

## 4. Discussion

PM$_{2.5}$ and ozone exposure are associated with subsequent day absences from elementary, middle, and high schools, even at low pollution levels. Furthermore, there is spatial and temporal variability of pollutant exposure in the SLCSD. Pollutant effects vary seasonally, with PM$_{2.5}$ having highest rate ratios with absences during the Fall, and ozone during the Spring. Using absence and pollutant exposure data at the individual school level we found that while pollution has similar effects on absenteeism, a higher level of baseline absences and of pollution on the more socioeconomically vulnerable West Side led to greater numbers of absences at these schools, despite smaller overall enrollment. This study considered morning temperature, influenza like illness hospital visits, and pollen counts as potential confounders and found that during the winter, both low temperatures and elevated PM$_{2.5}$ were associated with increased absences. The annual economic impact of pollution related absences in the SLCSD can provide a framework for quantifying the direct effects of air quality.

Although not explicitly examined in this study, sociodemographic factors are likely to have an effect on absences. These include, but are not limited to, location of residency, residence transiency, access to healthcare, transportation, parental education level, and nutritional options. These variables, along with other extraneous behavioral characteristics, such as whether students spend large amounts of time outside either through a sports activity or walking/cycling to and from school, may be substantial. As absences have established associations with negative educational outcomes, air quality may have longer-term socioeconomic ramifications beyond those shown in this study.

There are several limitations to this analysis. School location is a good approximation of residence for elementary and middle school students. However, as SLCSD enacts school choice initiatives, whereby older students are able to attend a non-neighborhood school, school location may be less reliable as an indirect indicator of air pollution exposure for high school students. Our analysis was restricted to ambient pollution exposure and could not consider either inhalational exposures in the home or indoor air quality in schools. However, epidemiologic studies have consistently found associations of ambient pollution and health outcomes, even for individuals, such as the elderly, who may spend very little time out of doors [8]. This study only considered students' current enrollment and did not attempt to consider historical residences or schools attended. The absences data set analyzed did not include individual student attendance information; therefore, these schoolwide estimates are not directly translatable to chronic absenteeism. The estimate of economic loss derived in this study is conservative, as we did not consider health care costs from childhood illnesses due to air pollution, transportation costs, and additional miscellaneous costs associated with school absence. Therefore, this analysis is not a comprehensive study, but rather, a starting point. Finally, the long-term effects of absenteeism on student success must be considered as a potential benefit of measures to improve air quality.

A recent report showed that the number of uninsured children is on the rise nationwide. Utah had the eight highest rate (72,000 or 7.4%), and second largest proportional increase (13,000 or 1.4%) in the number of uninsured children during the study period [48]. As children from disadvantaged communities are more likely to be uninsured and be exposed to higher levels of pollution, they may also be more vulnerable to chronic absenteeism and associated long-term effects.

## 5. Conclusions

Our work demonstrates that low-level exposure, even at levels compliant with NAAQS can affect absence rates. These affected all school levels and schools of different socioeconomic circumstances. Lagged low-level exposure showed higher rate ratios at elementary schools than at other school levels possibly because younger children's less-developed lungs are more susceptible to adverse health outcomes. An important strength of this study was the availability of precise, granular estimates of air pollution at each school in the study that we obtained with concerted dense modeling or local observations. Use of data and real-time analysis from such a granular network would allow detailed school recess guidance to prevent harm from poor air quality as current recess recommendations are based on data from a single regulatory monitoring site [49] that captures exposure at the level of individual school imprecisely. Since differences in $PM_{2.5}$ concentrations across the school district are commonly over 6 µg/m$^3$, our findings are important for school administrators as well for regulators.

Future work may examine absences on a per-student basis, and comparing trends in air quality to patterns in student absenteeism, we may be able to more tightly correlate air quality to student chronic absenteeism which would allow predictive monetization of the impact of air quality on a pupil's predicted lifetime earnings, career attainment, likelihood of facing incarceration, and other meaningful socioeconomic metrics.

**CRediT authorship contribution statement**

**Daniel L. Mendoza:** Conceptualization, Methodology, Software, Formal Analysis, Investigation, Data Curation, Writing – Original Draft, Writing – Review & Editing, Visualization, Project Administration, **Cheryl S. Pirozzi:** Methodology, Writing – Review & Editing, **Erik T. Crosman:** Methodology, Data Curation, Writing – Review & Editing, **Theodore G. Liou:** Methodology, Software, Formal Analysis, Data Curation, Writing – Review & Editing, **Yue Zhang:** Methodology, Software, Formal Analysis, Data Curation, Writing – Review & Editing, **Jessica J. Cleeves:** Methodology, Writing – Review & Editing,  **Stephen C. Bannister:** Methodology, Formal Analysis, Data Curation, Writing – Review & Editing,  **William R.L. Anderegg:** Methodology, Data Curation, Writing – Review & Editing,  **Robert Paine III:** Conceptualization, Methodology, Resources, Writing – Review & Editing, Supervision, Project Administration, Funding Acquisition.

**Declaration of Competing Interest**

The authors declare that they have no known competing financial interests or personal relationships that could have appeared to influence the work reported in this paper.

**Acknowledgements**

This study was supported by Grants: 5T32HL105321-07, R01 HL125520, Subpopulations and Intermediate Outcomes in COPD Study (SPIROMICS II), Prevention and Early Treatment of Acute Lung Injury (PETAL) Network, from the National Institutes of Health; LIOU13A0, LIOU14P0, LIOU14Y4, LIOU15Y4, from the Cystic Fibrosis Foundation; VA Merit Award, from the U.S. Department of Veterans Affairs (VA); Ben B and Iris M Margolis Family Foundation of Utah; and Claudia Ruth Goodrich Stevens Endowment Fund at the University of Utah. The content is solely the responsibility of the authors and does not necessarily represent the official views of the funding agencies. The authors would like to acknowledge: Philip Harrison, Utah Division of Air Quality; Wei Beadles, Utah Department of Health; Dr. Lexi Cunningham, Superintendent Salt Lake City School District, Yándary Chatwin, Salt Lake City School District; Senator Luz Escamilla and Representatives Angela Romero and Mark Wheatley Utah State Legislature; TRAX Project Team University of Utah; National Allergy Bureau; Duane J. Harris, Intermountain Allergy & Asthma Clinic in Draper, UT, John D. Horel, University of Utah.

**Appendix A**

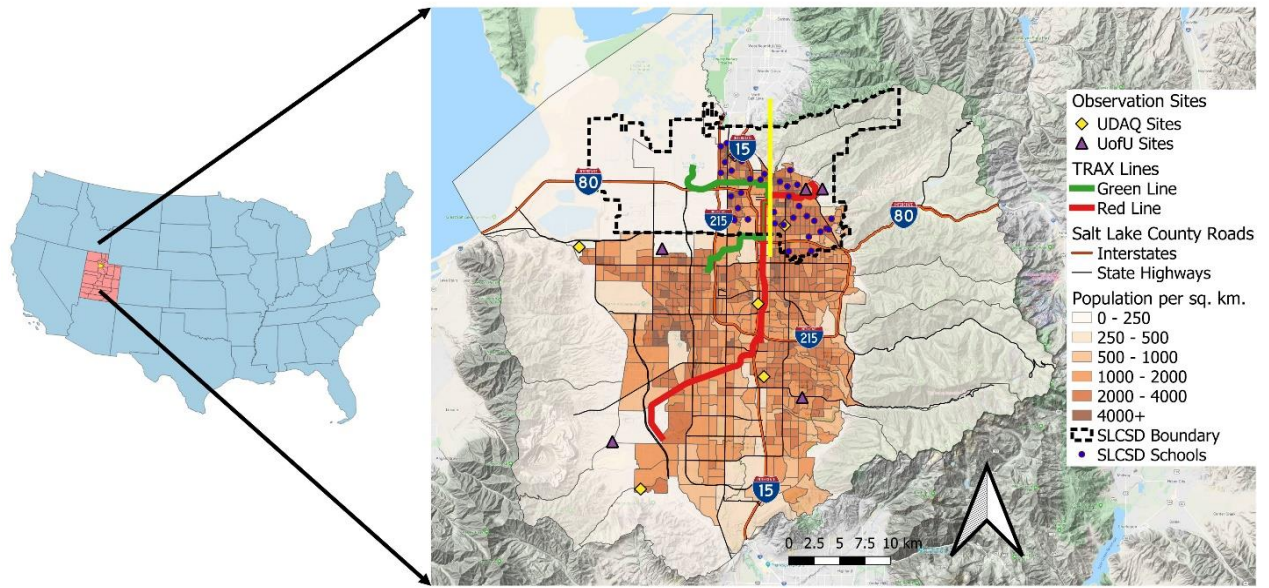

Fig. A.1. Map of Salt Lake City School District (SLCSD) schools (blue circles), Utah Division of Air Quality (yellow diamonds) and University of Utah (purple triangles) stationary observation sites, light rail routes (green and red lines), interstate highways (orange lines), state highways (black lines), and population density (color gradient) of Salt Lake County, Utah. The solid yellow vertical line divides the SLCSD (dashed black outline) into the west and east sides.

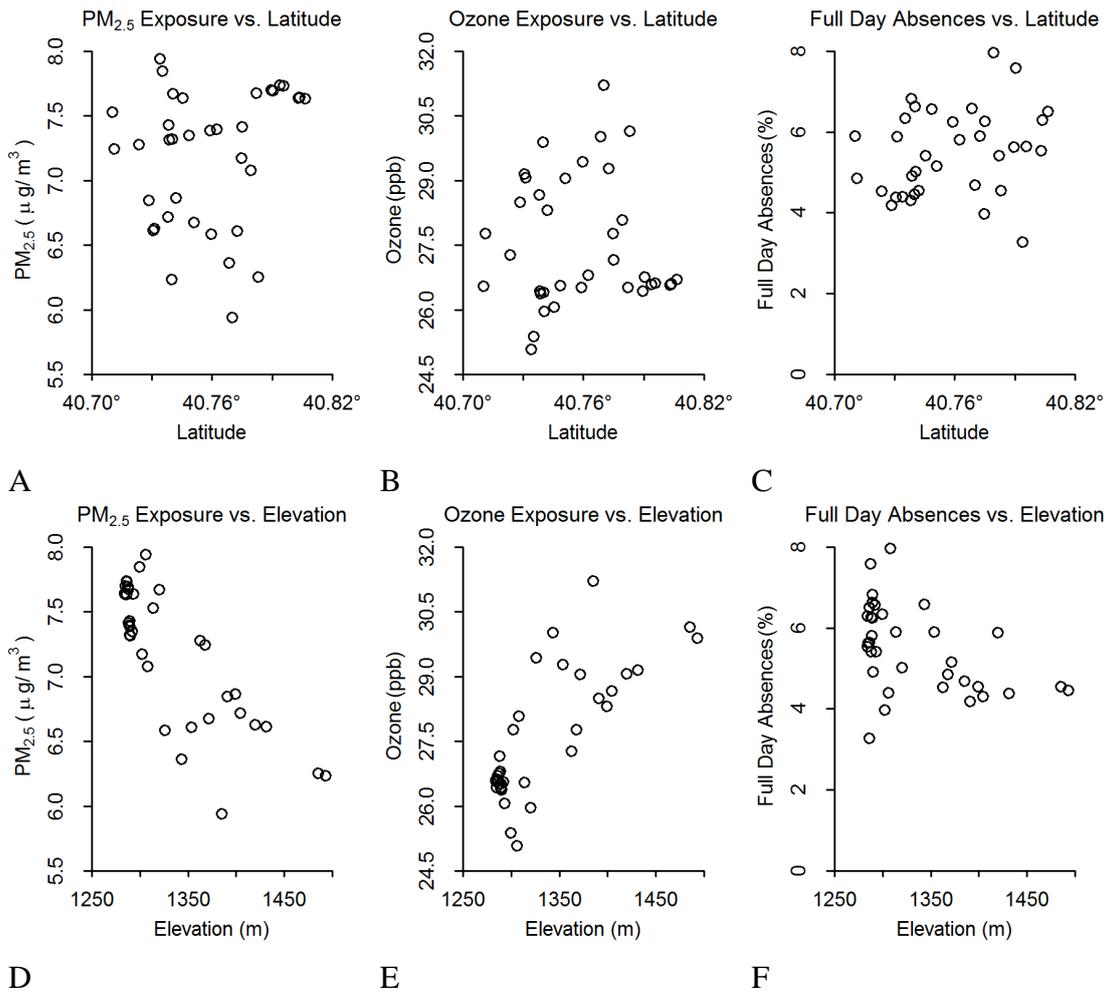

Fig. A.2. Comparison between average annual pollutant exposure and absences by geographical characteristics. Each circle represents a school location and its corresponding annual pollution exposure or absence rate: A) fine particulate matter ($PM_{2.5}$) vs. latitude, B) ozone vs. latitude, C) absences vs. latitude, D) $PM_{2.5}$ vs. elevation, E) ozone vs. elevation, and F) absences vs. elevation.

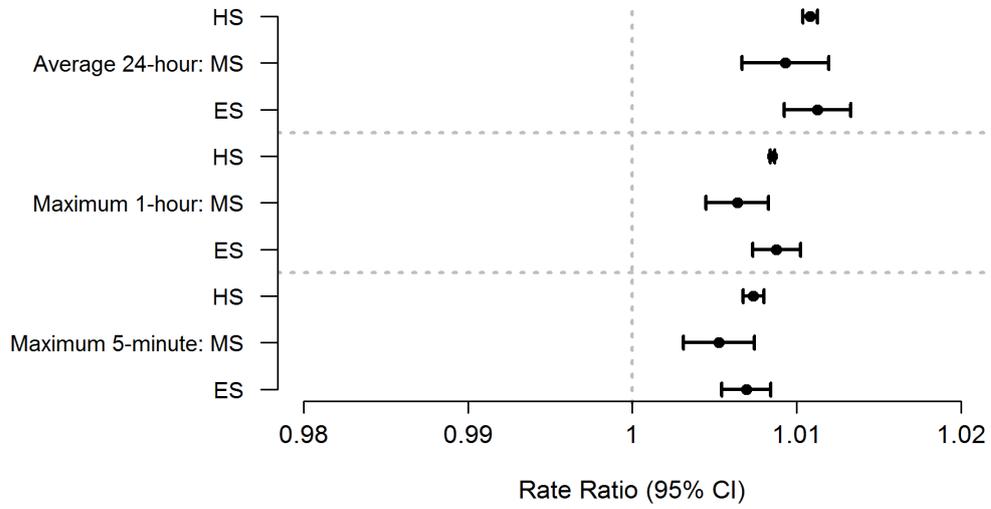

Fig. A.3. Rate ratios of Salt Lake City School District fall (September-November) absences associated with fine particulate matter (PM$_{2.5}$) exposure during school hours (7 AM - 3 PM).

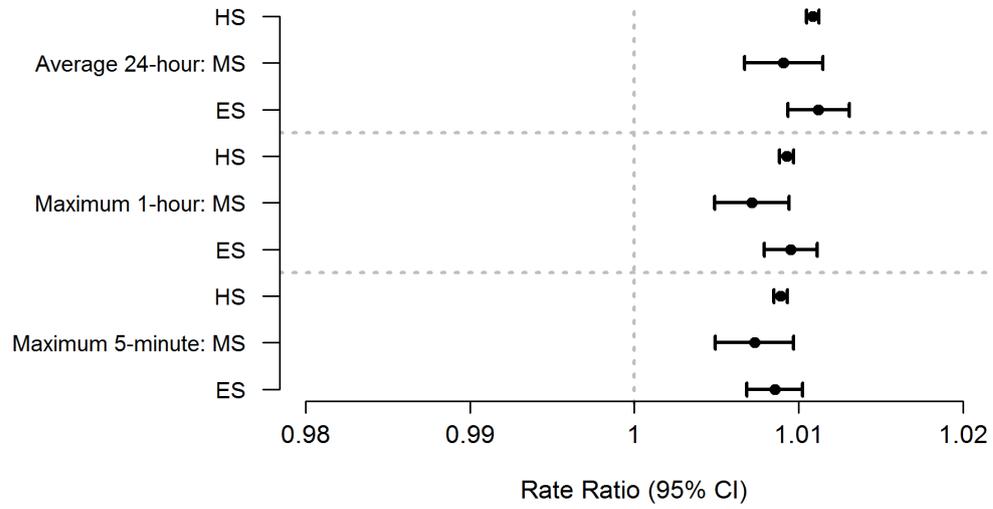

Fig. A.4. Rate ratios of Salt Lake City School District fall (September-November) absences associated with fine particulate matter (PM$_{2.5}$) exposure during recess hours (10 AM - 2 PM).

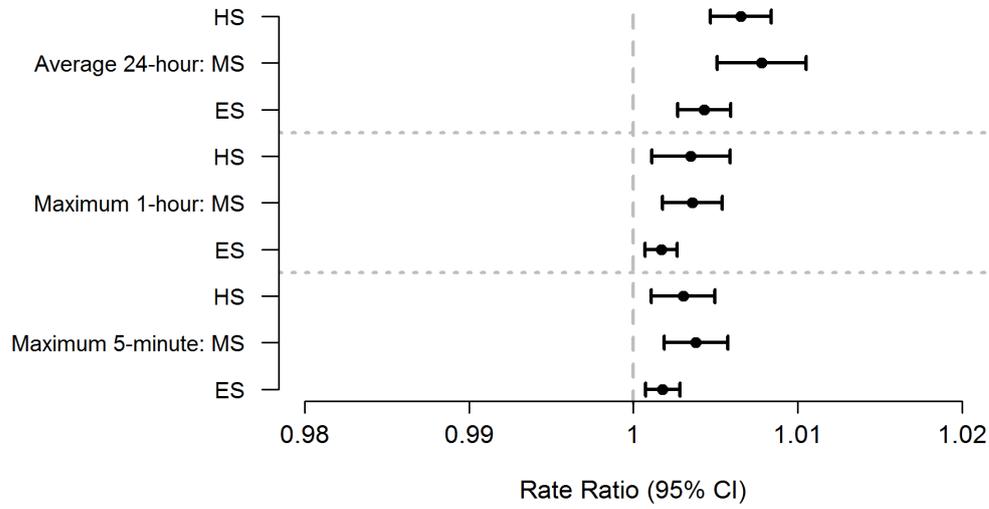

Fig. A.5. Rate ratios of Salt Lake City School District academic year (August-June) absences associated with previous day ozone exposure for all concentration levels.

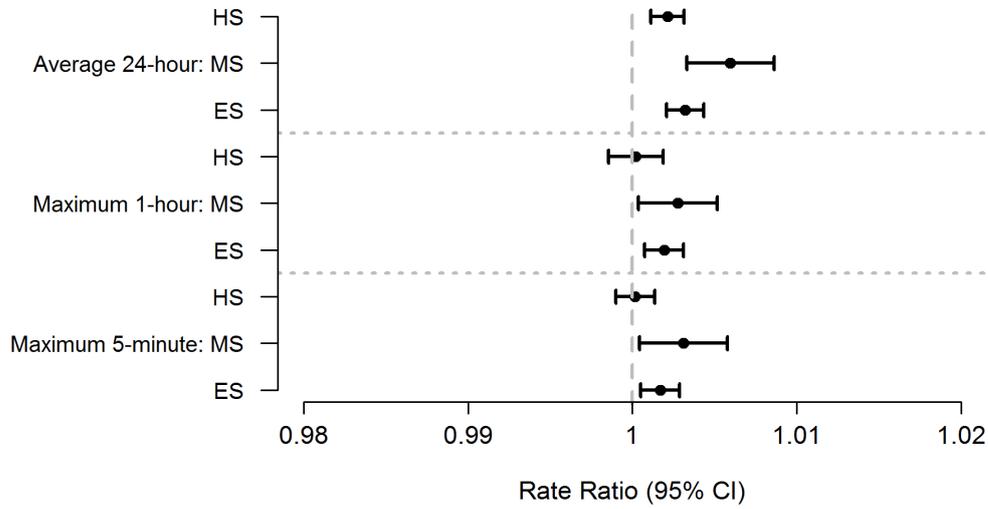

A

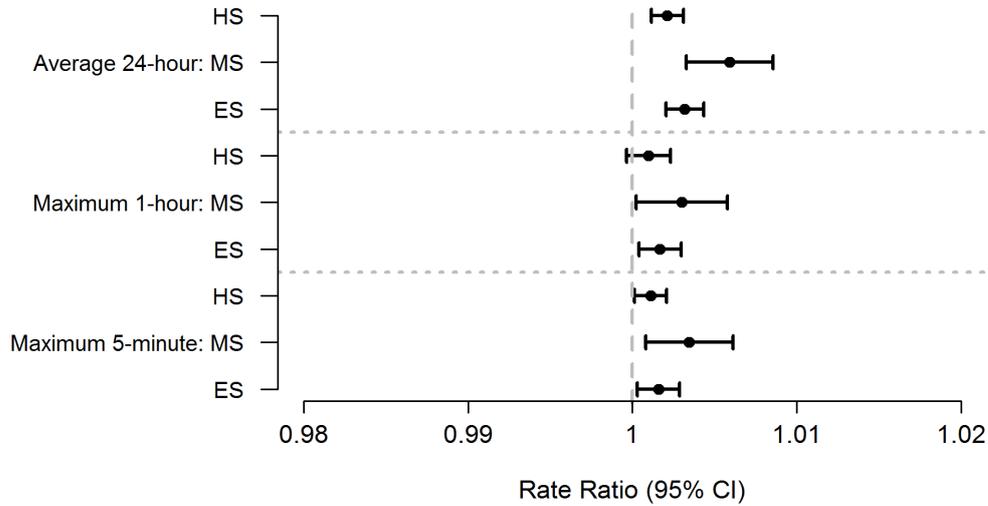

B

Fig. A.6. Rate ratios of Salt Lake City School District winter (December-February) absences associated with previous day ozone exposure for: A) all concentration levels and B) low (< 55 ppb) concentration levels.

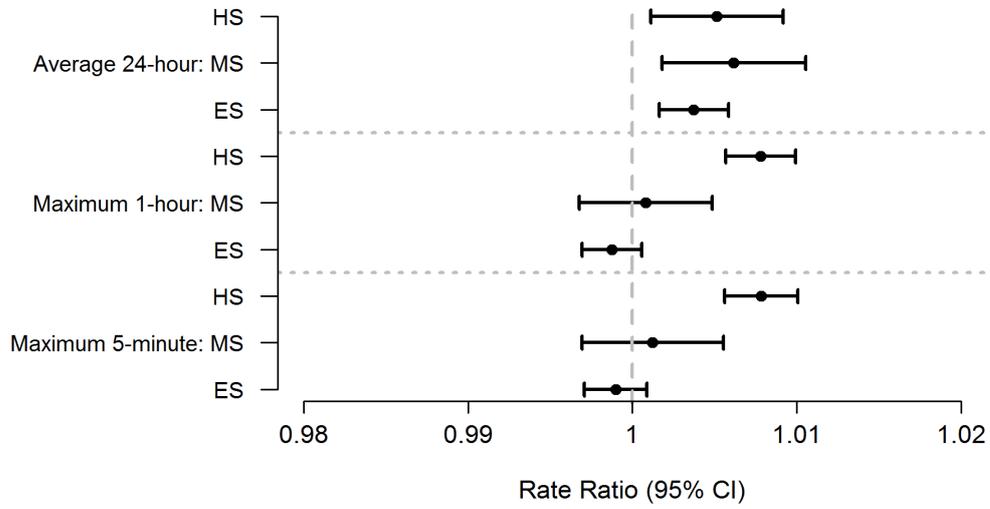

A

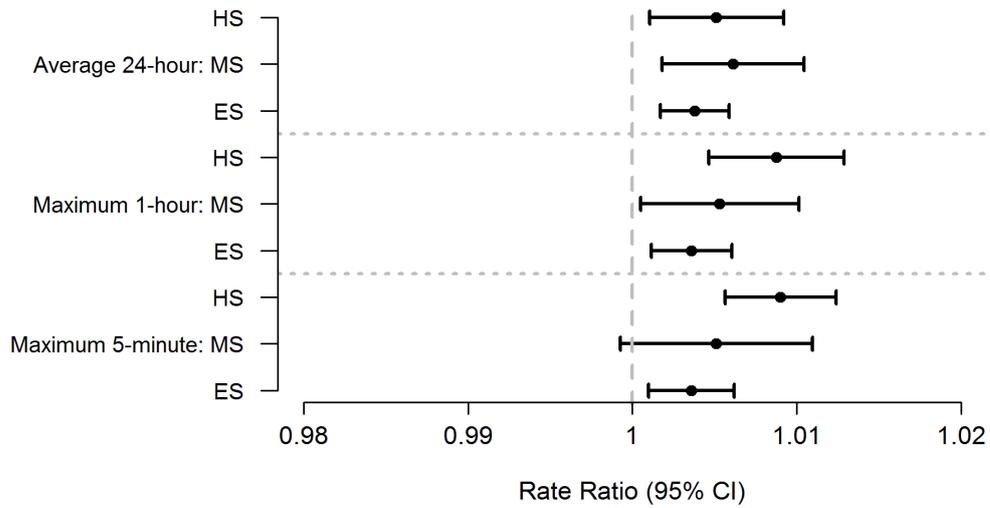

B

Fig. A.7. Rate ratios of Salt Lake City School District spring (March-May) absences associated with previous day ozone exposure for: A) all concentration levels and B) low (< 55 ppb) concentration levels.

Table A.1. Economic impact of school absences in the Salt Lake City School District (SLCSD) and potential benefits of pollution reduction disaggregated by school location within the SLCSD.

| School Level (N) | West (16) | East (20) | District (36) |
|---|---|---|---|
| **Title 1 Schools** | 15 | 4 | 19 |
| **Annual Enrollment: Mean (SD)** | 648 (512) | 619 (410) | 632 (451) |
| **Annual Attendance Days: Mean (SD)** | 116,615 (92,077) | 111,444 (73,711) | 113,742 (81,177) |
| **Annual Absences: Mean (SD)** | 6,337 (3,428) | 5,742 (3,488) | 6,007 (3,425) |
| **Absences (%): Mean (SD)** | 5.82 (1.08) | 5.40 (1.27) | 5.59 (1.19) |
| **Annual PM2.5 ($\mu g/m^3$): Mean (SD)** | 6.73 (0.08) | 6.43 (0.42) | 6.56 (0.35) |
| **Annual Ozone (ppb): Mean (SD)** | 28.17 (0.2) | 28.65 (0.88) | 28.44 (0.7) |
| **Absences Reduction: Mean (SD)** | 20 (10) | 16 (7) | 18 (9) |
| **Lost School Revenue ($): Mean (SD)** | 835 (420) | 652 (305) | 733 (367) |
| **Lost Wages ($): Mean (SD)** | 3,841 (1,933) | 2,997 (1,403) | 3,372 (1,689) |
| **Meal Costs ($): Mean (SD)** | 36 (19) | 28 (13) | 32 (16) |
| **Total Family Burden ($): Mean (SD)** | 3,877 (1,952) | 3,025 (1,416) | 3,404 (1,705) |
| **Economic Multiplier ($): Mean (SD)** | 9,602 (4,832) | 7,492 (3,509) | 8,430 (4,221) |
| **Total Economic Cost: Mean (SD)** | 14,314 (7,204) | 11,169 (5,230) | 12,567 (6,293) |